\documentclass[prl,aps,twocolumn, floatfix, superscriptaddress, showpacs]{revtex4}
\usepackage{amsmath,bm,graphicx}
\usepackage{ifthen}
\newboolean{colorFigures}
\setboolean{colorFigures}{true}

\begin{document}

\title{Growing condensate in two-dimensional turbulence}

\author{M. Chertkov}
\affiliation{Theoretical Division \& Center for Nonlinear Studies, LANL, Los Alamos, NM 87545, USA}
\author{C. Connaughton}
\affiliation{Theoretical Division \& Center for Nonlinear Studies, LANL, Los Alamos, NM 87545, USA}
\author{I. Kolokolov}
\affiliation{ Landau Institute for Theoretical Physics, Moscow, Kosygina 2, 119334, Russia}
\affiliation{Theoretical Division \& Center for Nonlinear Studies, LANL, Los Alamos, NM 87545, USA}
\author{V. Lebedev}
\affiliation{ Landau Institute for Theoretical Physics, Moscow, Kosygina 2, 119334, Russia}
\affiliation{Theoretical Division \& Center for Nonlinear Studies, LANL, Los Alamos, NM 87545, USA}

\date{\today}

\begin{abstract}
We report a numerical study, supplemented by phenomenological
explanations, of ``energy condensation'' in forced 2D turbulence in
a biperiodic box. Condensation is a finite size effect which occurs
after the standard inverse cascade reaches the size of the system.
It leads to emergence of a coherent vortex dipole. We show that the
time growth of the dipole is self-similar, and it contains most of
the injected energy, thus resulting in an energy spectrum which is
markedly steeper than the standard $k^{-5/3}$ one. Once the coherent
component is subtracted, however, the remaining fluctuations have a
spectrum close to $k^{-1}$. The fluctuations decay slowly as the
coherent part grows.
\end{abstract}
\maketitle

A big difference between 2D and
3D turbulence is the generation of large scale structures from small scale motions
\cite{frischBook,02KEL}. This occurs because, if pumped at intermediate scales,
the 2D Navier--Stokes equations favor energy transfer to larger scales
\cite{67Kra,68Lei,69Bat,06CHE}, a phenomenon known as an inverse cascade. Simulations
\cite{93SY,94SY} and experiments \cite{98PT,00JIN,05SHA} show that large scale
accumulation of energy is observed if conditions permit the energy to reach the system
size. In this letter, we study the ``condensate" emerging in the form of two coherent vortices in a
biperiodic box in 2D.
Let us begin by briefly reviewing the classical 2D turbulence theory of Kraichnan, Leith
and Batchelor (KLB) \cite{67Kra,68Lei,69Bat}. The essential difference with 3D turbulence
is the presence of a second inviscid invariant, in addition to energy, the enstrophy.
Stirring the 2D flow leads to emergence of two cascades. Enstrophy cascades from the
forcing scale, $l$, to smaller scales (direct cascade) while energy cascades from the
forcing scale to larger scales (inverse cascade). Viscosity dissipates enstrophy at the
Kolmogorov scale, $\eta$, which is much smaller than $l$ when the Reynolds number is
large. The energy cascade is blocked at a scale $\zeta$, $\zeta\gg l$, by a frictional
dissipation (usually due to friction between the fluid and substrate although other
mechanisms can be imagined) after a transient in time quasi-stationary regime. Then a
stationary KLB turbulence is established \cite{67Kra,68Lei,69Bat}. Applying Kolmogorov
phenomenology (see e.g. \cite{frischBook}) KLB predicts an energy spectrum scaling as
$k^{-3}$ in the direct cascade, and as $k^{-5/3}$ in the inverse cascade. Here $k$ is the
modulus of the wave-vector. The KLB spectra imply that velocity fluctuations at a scale
$r$, $\delta v_r$ scale as $\epsilon^{1/3}l^{-2/3} r$ and $(\epsilon r)^{1/3}$ in the
direct and inverse cascade ranges respectively. KLB theory is confirmed by simulations
\cite{00BCV,06BOF} and experiments \cite{05BRU, 02KEL}, where a sufficient range of scales was
available to form the cascades. If the frictional dissipation is weak so that $\zeta$
exceeds the system size $L$ then ultimately the ``condensate" regime emerges
\cite{67Kra,93SY} where the standard KLB does not apply.

One of the primary motivations for studying 2D turbulence is that it is structurally and
phenomenologically similar to quasi-geostrophic turbulence \cite{lesieurBook,71CHA} which
describes planetary atmospheres \cite{85NAS}. In addition, a recent resurgence in
theoretical interest in the condensate state was sparked by experimental \cite{05SHA} and
numerical \cite{04CLE} observations of large scale coherent vortices associated with
energy condensation in forced, bounded flows. In this letter, we report results of a set
of numerical experiments designed to give a clean, detailed study of this energy
condensation phenomenon in its own right.

\begin{figure*}
\begin{center}
\ifthenelse{\boolean{colorFigures}}
{
  \includegraphics{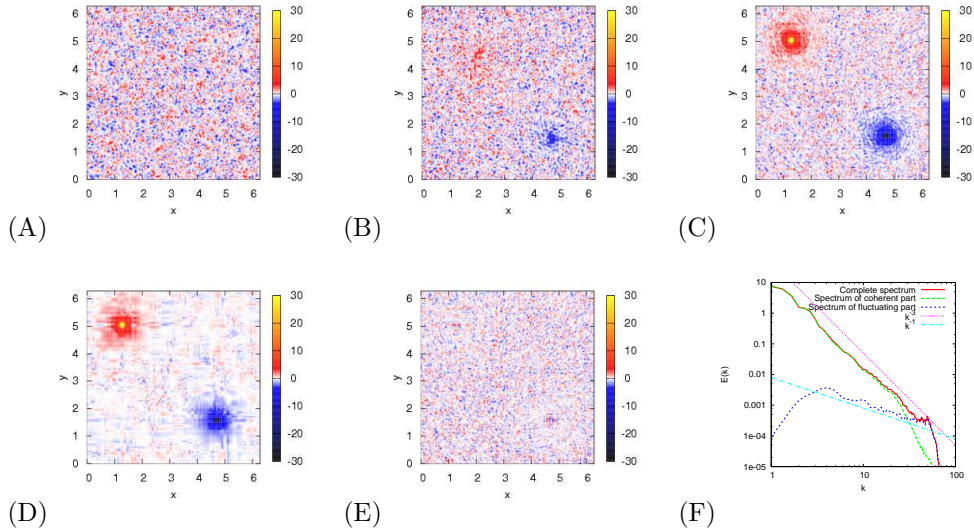}
}
{
  \includegraphics{panel1.mono.eps}
}
\end{center}
\caption{\label{fig-vorticitySnapshots} Vorticity snapshots at a succession of times :
(A) $t=100$, (B) $t=2000$, (C) $t=10000$. Decomposition of the $t=10000$ snapshot into
coherent, (D), and fluctuating parts, (E). Energy spectra of the full field, coherent
part and fluctuating part, (F).}
\end{figure*}

We solved the incompressible forced Navier--Stokes equations with
hyperviscous dissipation in 2D:
 \begin{eqnarray}
 \nonumber \partial_t {\bm u}+({\bm u \cdot \nabla}){\bm u}
 +{\bm \nabla}p &=& \nu\Delta^8{\bm u} +{\bm f} \\
 {\bm \nabla} \cdot {\bm u} &=&0. \label{NS}
 \end{eqnarray}
The domain is a doubly-periodic box of size, $L=2\pi$. The forcing,
${\bm f}$, injects velocity fluctuations and energy at an
intermediate scale $l$ with energy injection rate $\epsilon$. For
the simulations shown in Figs.~\ref{fig-vorticitySnapshots} and 
~\ref{fig-spectrumEvolution},
$l=2\pi/50$ and $\epsilon=0.004$. We use a standard pseudo-spectral
solver with full dealiasing. The resolution varied from $256^2$ to
$1024^2$. For developed condensate computations, the resolution was
only $256^2$ owing to the requirement of integration for tens of
thousands of forcing times, which was done using a 3rd order
Runge--Kutta integrator with integrating factors.  The timestep was
decreased as the condensate grows such that it satisfies $\Delta t <
c_0 \Delta x/ u_{\rm max}$, where $\Delta x$ is the grid spacing,
$u_{\rm max}$ is the maximum velocity and $c_0$ is conservatively
taken in the range $0.2$ -- $0.5$. Energy injection was done in a
spectral band using a stochastic additive force with fixed amplitude
and random phase. The correlation time is the numerical timestep.
Small scale dissipation was provided by $\Delta^8$ hyperviscosity
which is not expected to affect the large scale
behavior. There is evidence \cite{01DAN}, and our additional
simulations (not discussed here) are in support, that the presence
of large scale damping can considerably complicate the description
of the large scales.

The forcing described above is short correlated in time and characterized by the energy 
injection rate, $\epsilon$, and the forcing scale, $l$. The majority of 
the injected enstrophy cascades towards smaller
scales to be dissipated by viscosity. In our numerics, the closeness of $\eta$ to $l$
meant that only 40\% of the injected energy goes upscale from $l$ with the rest going
downscale with the enstrophy. We simulate the zero friction case to assure that
eventually the energy starts piling up at the system size scale. The direct cascade is
set up in a short time, $\tau_l\sim l^{2/3}/\epsilon^{1/3}$. The time, $\tau_*$, for the
inverse cascade to populate all scales up to $L$, is much longer. Based on Kolmogorov
arguments, $\tau_*\sim \epsilon^{-1/3} L^{2/3}$. In the simulations, $\tau_*\approx
1000$, in units where $\tau_l$ is about 1.

At $t>\tau_*$ we observed a condensate consisting of two big
vortices of the size $\sim L$ separated by a hyperbolic domain of
comparable size. Fig.~\ref{fig-vorticitySnapshots} (A),(B) and (C)
illustrates the phenomenon with a series of vorticity snapshots. The
condensate is formed to ensure that (a) the integral vorticity is
zero in accordance with zero integral vorticity injected by the
small-scale pumping, and (b) majority of energy brought by the
inverse cascade is accumulated at the largest scale, $L$. Two
identical vortices rotating in opposite directions satisfy these
conditions. Due to biperiodicity, Fig.~\ref{fig-vorticitySnapshots}
actually depicts the emergence of a vortex crystal. Such crystals have
been observed both numerically \cite{94SY} and experimentally
\cite{00JIN}. The vortices drift slowly over time but the square
symmetry of the crystal is preserved by this drift.

Evolution of the vortices is slow relative to the background fluctuations which permits a
separation of the flow into coherent and fluctuating components in the spirit 
of \cite{88FR, 99FSK}. The highest amplitude coefficients of the wavelet 
transformed vorticity are assigned to
the coherent component and the remainder to the incoherent component.  Inverse 
wavelet transforms are then taken.
This
decomposition is shown in Fig.~\ref{fig-vorticitySnapshots} (D) and (E).
Fig.~\ref{fig-vorticitySnapshots}(E) shows that the fluctuating part is almost
statistically homogeneous, whereas the coherent part is strongly inhomogeneous. Note that
the characteristic amplitude of the vorticity fluctuations is larger than the coherent
part of the vorticity over most of the domain. Ultimately we expect the
coherent flow to dominate the fluctuations everywhere but we have not reached this
regime.

\begin{figure}[t]
\begin{center}
\ifthenelse{\boolean{colorFigures}}
{
  \includegraphics{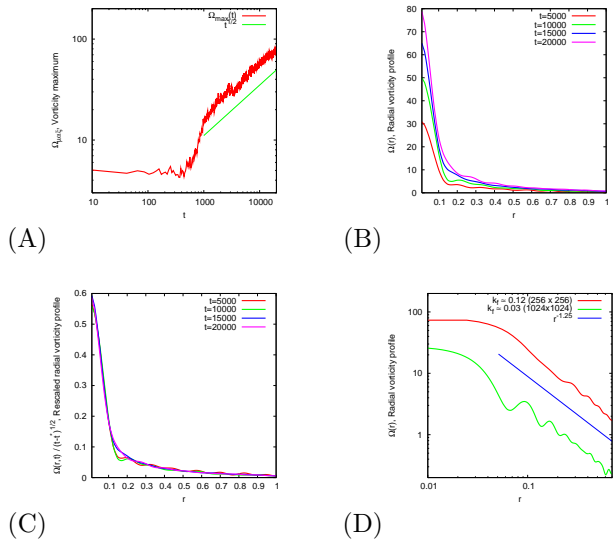}
}
{
  \includegraphics{panel2.mono.eps}
}
\end{center}
\caption{\label{fig-condensateSelfSimilarity} Self-similar growth of the condensate.
(A) Maximum vorticity as a function of time, (B) Angle--averaged vorticity,
$\Omega(r)$, as a function of distance, $r$, from the vortex center for 
successive times, (C) Same profiles rescaled by $\sqrt{t-t^*}$, (D) $\Omega(r)$
in the developed condensate regime for 256x256 and 1024x1024 simulations with
two different forcing scales.}
\end{figure}

As seen in Fig.~\ref{fig-condensateSelfSimilarity}(A), one
observes $\propto \sqrt{t}$ growth of the maximum value of the
coherent part of vorticity with time. Furthermore, simulations show
the global growth $\propto\sqrt t$ of the coherent velocity profile.
This global self-similarity is evident from
Figs.~\ref{fig-condensateSelfSimilarity}(B) and
\ref{fig-condensateSelfSimilarity}(C). The law $\propto\sqrt t$ is
naturally explained by the energy accumulation injected at the
constant rate, $\epsilon$, by forcing. In the hyperbolic region one
estimates the coherent velocity as $\sqrt{\epsilon t}$.

The mean velocity profile within the vortex is almost perfectly circular. To a good
precision higher order harmonics are suppressed relative to the zeroth order one. The
velocity profile deduced from the simulations fits, is $\propto r^{-\xi}$, where
$\xi\approx 0.25$, in the range, $L\gg r\gg l$, and thus the vortex core is roughly $l$.
This is illustrated in Fig.~\ref{fig-condensateSelfSimilarity}(D) showing the equivalent
vorticity profile, $\propto r^{-1.25}$. We plot $r^{-1.25}$ profiles for two different
forcing scales to check that the profile is insensitive to it.

\begin{figure}[t]
\begin{center}
\ifthenelse{\boolean{colorFigures}}
{
  \includegraphics[width=5.0cm]{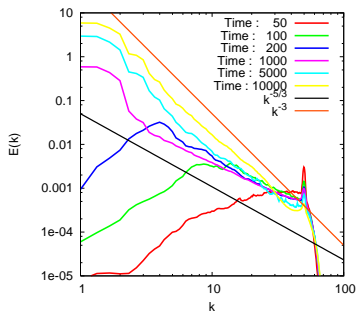}
}
{
  \includegraphics[width=5.0cm]{panel3.mono.eps}
}
\end{center}
\caption{\label{fig-spectrumEvolution} Time evolution of the
spectrum showing the transition from a standard $k^{-5/3}$ inverse
cascade to a $k^{-3}$ scaling at large scales. Spectra have been locally
averaged in time.}
\end{figure}

So far in this letter we have been discussing the spatio-temporal features of the
condensate. Complementarily, and following tradition developed in turbulent studies, one
may also analyze velocity spectra. Evolution of the spectra in time is shown in
Fig.~(\ref{fig-spectrumEvolution}), where one clearly see transition at $t_*$ from
$k^{-5/3}$ to scaling steeper than $k^{-5/3}$, that is numerically close to $k^{-3}$.
Similar statements were made before in Refs. \cite{94Bor,04TB}. We claim that this
spectrum is spurious and should not be taken as evidence of a cascade in the KLB sense.
The coherent part of the flow has almost no fluctuations and, if it is
removed, the steeper than $k^{-5/3}$ scaling disappears entirely. By contrast,
the $k^{-3}$ enstrophy cascade of KLB involves fluctuating vortices across
many scales.

\begin{figure}
\begin{center}
\ifthenelse{\boolean{colorFigures}}
{
\includegraphics{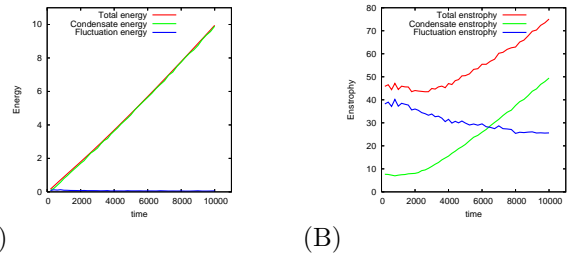}
}
{
\includegraphics{panel4.mono.eps}
}
\end{center}
\caption{\label{fig-fluctuationDecay} (A) Time evolution of the energy
contained in the condensate and background fluctuations (B) enstrophy.}
\end{figure}

The spectrum of the fluctuations is shown in Fig.~\ref{fig-vorticitySnapshots}(F),
it is close to $k^{-1}$. There is some evidence that shortly after $\tau_*$, the
fluctuations retain a $k^{-5/3}$ spectrum as was observed in \cite{94Bor,04TB}. However,
this regime is transient and does not persist for long after the condensate emerges.
Larger resolution will be required to settle this question unambiguously. Fluctuations
play a relatively minor role in the overall energy balance with the majority of the
energy absorbed into the condensate as shown in Fig.~(\ref{fig-fluctuationDecay})(A).
They contain more of the enstrophy as shown in Fig.~(\ref{fig-fluctuationDecay})(B), but
they decay in amplitude as the condensate grows so that the flow becomes more and more
coherent as time passes. The data suggest that decay of the amplitude of the background
fluctuations is logarithmic, or very weakly power law.

We now present an attempt to describe phenomenologically, the
universal nature of the asymptotic condensate state. Consider an
individual vortex at $t\gg \tau_*$. It has a core of radius
$\sim l$ and its spatial extent is estimated by the system
size, $L$. In discussing the spatial structure of the vortex, for example its
mean vorticity profile, $\Omega=\langle\nabla\times{\bm u}\rangle$,
we will track its dependence on the distance, $r$, from the center
of the vortex, $\Omega(r)$. Once the almost circular vortex emerges,
it keeps sucking energy from the turbulent background which can be
approximately described by an inhomogeneous eddy diffusivity,
$D(r)$. Another large scale characteristic affected by the
eddy-diffusivity is the fluctuation enstrophy, $H(r)=\langle(\nabla\times{\bm
u}-\Omega)^2\rangle$. The focused regime is adiabatic so that the
equations governing the quasi-stationary radially symmetric
distribution of $\Omega$ and $H$ on the top of the turbulent
background are the eddy-diffusivity equations
\begin{equation}
\partial_r rD\partial_r\Omega=0,\quad \partial_r r D\partial_r H=0.
\label{eddy_diff}
\end{equation}
Naturally, $D$ can be expressed in terms of the typical Lyapunov
exponent,  $\lambda$, $D\sim r^2\lambda$. In  homogeneneous
turbulence $\lambda$ would be self-consistently estimated as
$\sim\sqrt{H}$. However present situation is inhomogeneous, with a strong,
 $\sim\Omega$, shear. Mixing in the
presence of  strong shear was discussed in \cite{05CKLT}. It was shown
that the dependence of the effective Lyapunov exponent on the mean
shear, $\sim \Omega$, and the background enstrophy, $H$, can be
estimated as
\begin{equation}
 \lambda\sim \left\{\begin{array}{cc}
 H^{1/6}\Omega^{2/3}, & \tau_H\lambda\ll 1;\\
 H^{1/4}\Omega^{1/2}, & \tau_H\lambda\gg 1.\end{array}\right.
 \label{Lambda}
\end{equation}
Here $\tau_H$ is the correlation time of the background vorticity
fluctuations. The actual nonparametric regime we are interested in
is $\tau_H\lambda\sim 1$. Thus keeping the two asymptotics in
Eq.~(\ref{Lambda}) will, in fact, give upper and lower bounds.
Returning to Eqs.~(\ref{eddy_diff}) one notes that the physically
meaningful solution for $\Omega$ corresponds to a flux state zero
mode of the eddy-diffusivity equation, describing accumulation of
the mean vorticity (and the mean energy) inside the vortex,
$Dr\partial_r\Omega=\mbox{const (wrt } r)$. On the contrary, the
physically meaningful solution of the eddy-diffusivity equation for
$H$ is the one corresponding to a zero spatial flux, $H=\mbox{const
(wrt } r)$. Note, that this spatially homogeneous distribution of
$H$ is in agreement with the results of simulations. Combining all
these estimations with the global energy conservation one arrives at
the following bounds
\begin{equation}
\Omega\sim \frac{\sqrt{\epsilon t}}{r}\left\{\begin{array}{cc}
 (L/r)^{1/5}, & \tau_H\lambda\ll 1;\\
 (L/r)^{1/3}, & \tau_H\lambda\gg 1.\end{array}\right.
 \label{Omega_profile}
\end{equation}
These estimates for the mean vorticity profile are fully
consistent, in describing both the overall temporal dynamics and the
exponent of the mean vorticity profile, with the aforementioned
numerical simulations, shown in Fig.~2:  $1/5<\xi\approx
0.25<1/3$. The corresponding estimate for the spatially homogeneous
enstrophy, also expressing direct enstrophy balance at the pumping
scale, is $H\sim \varepsilon/(l^2\lambda)$. This formula, combined
with Eq.(\ref{Lambda}) for the Lyapunov exponent, predicts a slow
algebraic decay of the background enstrophy in time. This is again
consistent with simulations.

Finally, the spatial homogeneity of $H$ suggests that majority of
the injected enstrophy cascades to smaller scales, $r\ll l$.
However, a subdominant portion will also penetrate to the larger
scales,  e.g. resulting in the $k^{-1}$ spectrum observed in the
simulations, see Fig.~3. An explanation of the $k^{-1}$ spectrum
observed at the scales larger than $l$ after subtraction of the
coherent component of the flow is as follows. In a range of scales
smaller than $L$ vorticity fluctuations are advected {\it
passively}. Passive scalar theory, developed in \cite{99BFLL},
predicts an $\sim 1/r^2$ decay for the pair correlation of a scalar
at the scales larger than injection scale in two dimensions and for
non-zero value of the Corrsin invariant,  which is the integral of
the pair correlation function of the pumping. However, vorticity is
a curl of velocity injection and thus the vorticity is injected at
$l$ with zero value of the Corrsin invariant. This leads to the
localized, $\sim\delta({\bm r})$, expression for the pair
correlation function of vorticity, which in turns translates into
the observed $k^{-1}$ spectrum. Notice that similar explanation for
this scaling,  referred to as passive inverse energy
cascade, was reported in \cite{00NL}.

To conclude, we performed numerical simulations of energy
condensation in forced 2D turbulence.
We split the flow into coherent and fluctuating parts, observed the
power-like shape of the coherent vortices and self-similar growth in
time of the coherent flow. The coherent structure is responsible for
the spurious $k^{-3}$ energy spectrum observed in previous numerical
experiments. The fluctuations have an energy spectrum of $k^{-1}$
and they diminish in importance as the condensate grows. We also
presented phenomenological description of the simulations.

We wish to thank A. Celani, E. Lunasin and L. Smith for advice on numerics and R. Ecke,
G. Eyink, G.  Falkovich and M. Shats for helpful discussions. This work was carried out
under the auspices of the National Nuclear Security Administration of the U.S. Department
of Energy at Los Alamos National Laboratory under Contract No. DE-AC52-06NA25396.


\end{document}